\RequirePackage{fix-cm}
\documentclass[pdftex,twocolumn,epjc3]{svjour3}          % twocolumn
\usepackage{color}
\RequirePackage[T1]{fontenc}
\RequirePackage{amsmath} 
\RequirePackage{flushend}
\usepackage{colordvi}
\usepackage{graphicx,amssymb,epsfig,amsmath}
\usepackage{epstopdf}
\RequirePackage{latexsym}

\newcommand{\be}{\begin{equation}}
\newcommand{\ee}{\end{equation}}
\newcommand{\bea}{\begin{eqnarray}}
\newcommand{\eea}{\end{eqnarray}}
\newcommand{\te[1]}{$\times10^{#1}$}

\journalname{Eur. Phys. J. D}
\begin{document}
\title{Quantum dynamics of few dipolar bosons in a double-well potential}

\author{Rhombik Roy \thanksref{e1,addr1}
        \and
        Barnali Chakrabarti \thanksref{e2,addr1,addr2}
        \and
        Andrea Trombettoni \thanksref{e3,addr3}
}

\thankstext{e1}{e-mail: rhombik.rs@presiuniv.ac.in}
\thankstext{e2}{e-mail: barnali.physics@presiuniv.ac.in}
\thankstext{e3}{e-mail: andreatr@sissa.it}

\institute{Department of Physics, Presidency University, 86/1   College Street, Kolkata 700073, India \label{addr1}
           \and
           The Abdus Salam International Center for Theoretical Physics, I34100 Trieste, Italy \label{addr2}
           \and
           CNR-IOM DEMOCRITOS Simulation Centre and SISSA, Via Bonomea 265, I-34136 Trieste, Italy\label{addr3}
}

\date{\today}
\abstractdc{
  We study the few-body dynamics of
  dipolar bosons in one-dimensional double-wells.
  Increasing the interaction strength, by investigating one-body observables  
  we study in the considered few-body systems tunneling oscillations,
  self-trapping and the regime exhibting an equilibrating behaviour.
  The corresponding two-body correlation dynamics exhibits a strong
  interplay between the interatomic correlation due to non-local nature
  of the repulsion and the inter-well coherence.
  We also study the link between the correlation dynamics and
  the occupation of natural orbitals of the one-body density matrix.
}
%\keywordsdc {Tunneling dynamics, Dipolar gases}
%\pacs {05.45.-a, 05.45.Mt, 05.30.Jp, 05.10.-a}
\maketitle

\section{Introduction} \label{intro}
The study of the tunneling quantum dynamics of ultracold atoms in double-well potentials is a major tool to investigate
macroscopic coherent effects and many-body properties~\cite{5}. Given the very high level of manipulation and control of quantum gases,
the direct observations of tunneling and nonlinear self-trapping is experimentally possible for ultracold
bosons~\cite{andrea1,prl95,natphy1,andrea2,Levy,prl98,nature455,LeB,contact4,Xhani19} and fermions~\cite{Liscience,Lidiss,Luick,Kwon}
both for double- and multi-well potentials.

When in the uncoupled wells the system can be described by macroscopic wavefunctions, one has that the superfluid dynamics
in the double-well potential realizes the so-called Bosonic Josepshon junction (BJJ). For a Bose-Einstein condensate in a double-well potential,
interactions between the particles plays a crucial role. When the initial population imbalance of the two wells is below a critical value,
Josephson oscillations are predicted~\cite{5,zapata}. These two different dynamical regimes, Josephson oscillations and nonlinear self-trapping, for a BEC in a double-well potential
are commonly described at the mean-field level~\cite{5,rag,st,ab,cat}.
To study effects of quantum fluctuations and collapse-revival,
one may resort to quantum two-mode models~\cite{Jav1997,Raghavan1999,Franzosi2000,smerzi2000,bosehubbard2,Giovanazzi,DellAnna,Bila}.
Beyond mean-field effects are very important for coupled one-dimensional systems~\cite{Jo,gritsev,Aga,aPolo18,Pigneur18,Polo}.

The solution of the time-dependent many-boson\\
 Schr\"odinger equation sheds light to several intriguing
  features, including the
role played by natural orbitals (i.e., the eigenstates of one-body density matrix) corresponding to the eigenvalues smaller than
the largest one~\cite{25,36,s40,fischer15}. The many-body tunneling dynamics in one dimensional double-well trap
with contact interactions has been also studied considering the
full crossover from weak interaction to the fermionization limit, and also attractive interactions
with~\cite{prl100,contact2,manybody1,contact3,Sudip2019}.

The experimental realization of dipolar gases, achieved with ultracold dipolar atoms of chromium~\cite{chromium1,chromium2},
dysprosium~\cite{dysprosium} and erbium atom~\cite{erbium}, further enlarged the range of phenomena that can be studied~\cite{30,31},
such as coupled one-dimensional systems with tunable dipolar interaction~\cite{tang19}.
The non-local nature of the dipolar interaction motivated several studies having as a goal the comparison
of results for the quantum dynamics with non-local interactions with the corresponding short-range findings
and with recent results for quantum systems with long-range
couplings~\cite{vodola14,gong16,lepori16,celardo16,defenu16,lepori17,igloi18,blass18,defenu18,lerose19}. 

The ground state properties of ultracold trapped bosons with dipolar interactions have been studied extensively both at mean-field level
and in the one-dimensional limit~\cite{39,Sinha,Citro,35,38,Cinti,DeP,abad1,abad2,abad3} (see more refs. in~\cite{30,31}).  
Properties of dipolar bosons in a double well potential were studied
by mean-field~\cite{blume} and by the quntum two-mode model~\cite{mazzarella}.
The many-body dynamics of bosons with non-local interaction has been recently studied in~\cite{halder,Sudip2019conf}, addressing the effect
of a finite-range interaction on density oscillation, collapse and self-trapping and focusing on the
comparison between the many-body and mean-field properties. The tunneling dynamics is studied for an interaction potential
modeled as $W(r)= \frac{\lambda_0}{\sqrt{(r/D)^{2n}+1}}$ with tunable strength $\lambda_0$ and half width $D$. 

In the present article we would like to address
the quantum dynamics of few dipolar bosons in a
one-\\dimensional double-well. The few body systems
we are going to study have the advantage that it is possible
to investigate in detail the full one-body density matrix and correlation
functions. In this way the effect of the non-local nature of the interaction
and of the one-dimensional nature of the potential on the quantum dynamics
can be investigated in details, and the dynamical regimes determined
by them can be visualized and understood. This will give insights on the more
complicated problem of many-body dynamics of one-dimensional dipolar
atoms in double-well dynamics and their quantum tunneling effects
to characterize what are the differences and anologes with the usual
mean-field findings.
Moreover, our results will be lead us to to
make a qualitative link between the observed dynamics of quantum correlations with the interatomic and inter-well correlations. The observed tunneling
dynamics is also connected with the occupation of the natural orbitals.

The non-local, dipolar interaction is modeled as
\begin{equation}
\hspace{18ex} W(r) = \frac{g_d}{r^3+\alpha_0}
\label{interp}
\end{equation}
 ($r=|x_i-x_j|$ being the distance between two particles located
in $x_i$ and $x_j$),where $g_d$ is the strength of dipolar interaction and $\alpha_0$ is a cutoff parameter to avoid
the divergence at $x_i=x_j$ (see the corresponding discussion in \cite{Sinha}). $g_d$ is determined by the scattering
length and transverse confinement~\cite{Olshanii:98}. In the follwing we keep $g_d > 0$, corresponds to repulsive dipolar bosons.
We prepare the initial state with complete population imbalance in the asymmetric double-well, with all bosons stayng in the right well.
The asymmetric double-well is modeled for simplicity as a superposition of three terms: a harmonic oscillator, a Gaussian with a central barrier
and a linear external potential~\cite{prl100}. In dimensionless units $V$ reads
\begin{equation}
\hspace{12ex} V(x)=\frac{1}{2}x^2 + V_0 \frac{e^{-\frac{x^2}{\sigma^2}}}{\sqrt{\pi} \sigma}- dx,
\label{potential}
\end{equation} 
with the parameter $d$ is initally kept sufficiently large to make the right well energetically favorable.
The initial state is prepared with practically all atoms in the right well. To study the dynamics in the symmetric double-well,
$d$ is then instantaneously ramped down to zero at $t=0$.
We solve the time-dependent Schr\"odinger equation for $N=6$ dipolar bosons.
The tunneling dynamics is monitored by the one-body density, the population in the right well, the population imbalance and the
two-body density for various choices of $g_d$. At $g_d=0$, we observe pure Rabi oscillation~\cite{Leggett}.
With very weak interactions, the Rabi oscillation is modified in amplitude. With larger $g_d$, self-trapping is observed. When $g_d$
is sufficiently strong, a new regime sets in and the bosons appear to equilibrate in two wells. These four kind of dynamics
are further linked with intricated interplay interatomic correlation and inter-well coherence.
We observe that, for the considered number of particles, the
many-body state occupies many natural orbitals in this one-dimensional setup also when $g_d$ is small. A marked occupation of different natural
orbitals is observed for very high value of $g_d$, where a vanishig population imbalance is observed for large times.
%We show that the dynamics of bosonic Josephson junction for the dipolar bosons is much richer than what is known for contact interaction. \\

The structure of the paper is as follows. In Sec.~\ref{method} we discuss the setup and give a brief introduction to the used
numerical method. In Sec.~\ref{result} we present our results, while summary and conclusions are given in Sec.~\ref{conclusion}.

%================================================================================================================================================================
\section{The model} \label{method}
The time-dependent many-body Schr\"odinger equation for $N$ interacting bosons is given by

\begin{equation}
\hspace{22ex} i\partial_t \vert \Psi \rangle = \hat{H} \vert \Psi \rangle
\label{TDSE}
\end{equation}

(with $\hbar=1$). Here, the Hamiltonian $\hat{H}$ is given by

\begin{equation}
\hspace{3ex}  \hat{H}(x_1,x_2,...,x_N)= \sum_{i=1}^{N} \hat{h}(x_i) + %\Theta(t)
  \sum_{i<j=1} \hat{W}(x_i-x_j).
\label{Hamiltonian}
\end{equation}
$\hat{h}(x)=\hat{T}(x)+\hat{V}(x)$ is the one-body term in the Hamiltonian, with the potential
$V$ given by Eq. (\ref{potential}) with $d=0$ during the dynamics and chosen (for $t<0$) to be large to determine the initial state as described
in the Introduction. $W$ is the interparticle potential, given
in Eq. (\ref{interp}).
The total Hamiltonian $\hat{H}$ is written
in dimensionless units, as obtained by dividing the dimensionful Hamiltonian by $\frac{\hbar^2}{mL^2}$ ($m$ is the mass of the bosons and $L$ is
an appropriately chosen length scale).

In the following Eq.(\ref{Hamiltonian}) is solved by the numerical many-body method called multi-configuration time-dependent Hartree method for bosons
(MCTDHB)~\cite{Streltsov2007,Ofir2008} implemented in the MCTDH-X software~\cite{x,y,axelN2,axelN3}. The MCTDHB has been extensively used in
different trapping potentials and interactions~\cite{budha1,budha2,budha3,MCTDHB_OCT,MCTDHB_Shapiro,cao1,cao2,NJP2015Schmelcher,axelN6,Axel2017,NJP2017Camille,NJP2017Schmelcher,rhombik,rhombik2,rhombik3}. MCTDH-X has been
verified against experimental predictions~\cite{axelprx} and is reviewed in~\cite{Ofir2019Rev}.

The many-body wavefunction in a complete set of time-dependent permanents, distributing $N$ bosons in $M$ time-dependent single particle orbitals. Thus the ansatz for the many-body wavefunction is
\begin{equation}
\hspace{18ex}   \vert \Psi(t)\rangle = \sum_{\bar{n}}^{} C_{\bar{n}}(t)\vert \bar{n};t\rangle,
\label{many_body_wf}
\end{equation}
with
\begin{equation}
\hspace{12ex} \vert \bar{n};t\rangle  = \prod_{i=1}^{M}
\left( \frac{ \left( b_{i}^{\dagger}(t) \right)^{n_{i}} } {\sqrt{n_{i}!}} \right) \vert vac \rangle.
\label{many_body_wf_2}
\end{equation}
In Eq. (\ref{many_body_wf}) the summation runs over all possible configurations $$N_{conf} = \left(\begin{array}{c} N+M-1 \\ N \end{array}\right).$$
The $\vert \bar{n};t\rangle= \vert n_1,....,n_M;t\rangle$ (with $\sum_i n_i\equiv N$) are the time-dependent permanents,
the operators $b_{i}^{\dagger} (t)$ create a boson in the \textit{$i$th} single particle state $\phi_i(x,t)$, and
$\vert vac \rangle$ is the vacuum.
The bosonic annihilation and corresponding creation operators obey the canonical commutation relation
$[b_{k},b_{j}^{\dagger}(t)]=\delta_{kj}$ at any time. It is important to emphasize that in the ansatz (\ref{many_body_wf})
both the expansion coefficients $\left\lbrace C_{\vec{n}}(t); \sum_i n_i=N \right\rbrace $ and the orbitals
$ \left\lbrace \phi_i(x_i,t)\right\rbrace_{i=1}^M $ that build up the permanents $\vert \bar{n};t\rangle$ are time-dependent and
fully variationally optimized quantities.

In the limit of $M \rightarrow \infty$, the expansion (\ref{many_body_wf}) is exact. However, we limit the size of the
Hilbert space during our computation requiring proper convergence. As permanents are time-dependent,
a given degree of accuracy is reached with a shorter expansion compared to time-independent basis.
To solve the time-dependent wavefunction $\Psi(t)$, we utilize the time-dependent variational principle~\cite{LF1,LF2}.
We substitute the many-body ansatz into the functional action of the time-dependent Schr\"odinger equation and require the
stationarity of the functional action with respect to $\{C_{\bar{n}}(t)\}$ and $ \left\lbrace \phi_n(\bar{r},t)\right\rbrace$ which
lead to the working equations of the MCTDHB method~\cite{Ofir2008,axelN3}.   

\begin{figure}  %%%fig1 imbalance3d
	\begin{center}
		\includegraphics[height=.50\textwidth,angle=0]{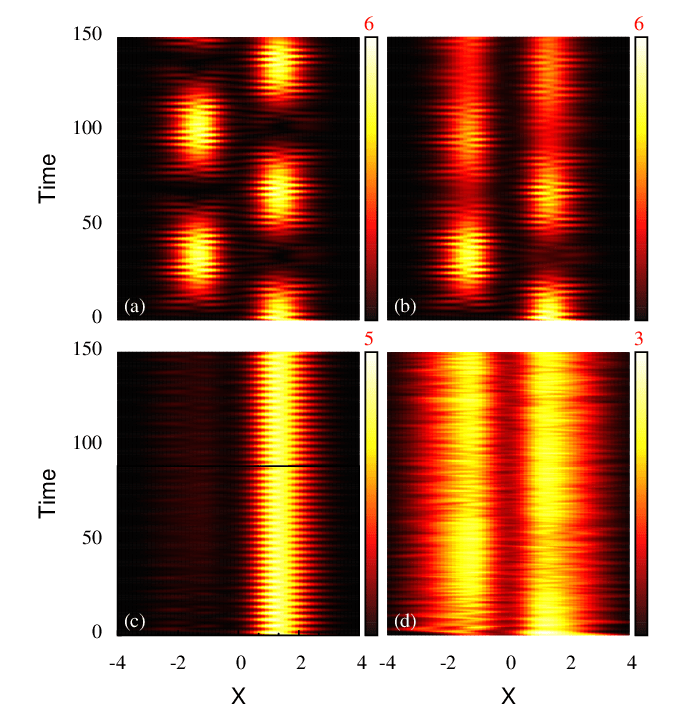}
	\end{center}
	\caption{Tunneling dynamics of $N=6$ interacting dipolar bosons in the double-well potential for different interaction strengths $g_d$.
          The population of the left and right well as a function of time is presented.
          (a) $g_d = 0$, Rabi oscillations. (b) $g_g= 0.002$, deformed Rabi oscillations. (c) $g_d= 0.02$, self-trapping in right well.
          (d) $g_d= 0.2$, equilibration in both wells. }
	\label{Fig1}
\end{figure}

\begin{figure}  %%%fig2 imbalance
	\begin{center}
		\includegraphics[height=.50\textwidth,angle=-90]{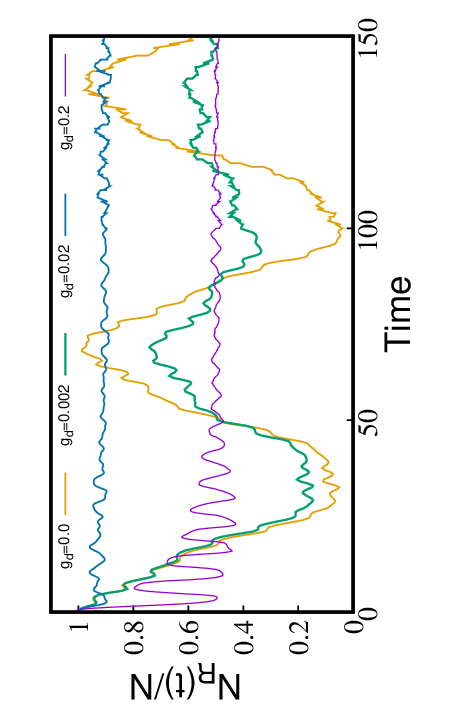}
	\end{center}
	\caption{Time evolution of the relative population in the right-hand well ($\frac{N_R(t)}{N}$) for $N=6$ dipolar bosons in double-well potential for different interaction strengths $g_d$.}
	\label{Fig2}
\end{figure}

%================================================================================================================================================================
\section{Results }\label{result}
To prepare the initial state, we propagate the equations of motion in imaginary time starting for an initial guess, so to have the
ground state of $N=6$ interacting bosons
in the right well of the asymmetric double-well $V(x)$ given by Eq. (\ref{potential}) at $t<0$. Chosen values of the parameters are
$V_0=8.0$ and $\sigma =0.7$. To study the dynamics in the symmetric double-well we ramp down $d \rightarrow 0$ at $t=0$.
We consider increasing values of $g_d$. We choose $g_d= 0,0.002,0.02,0.2$ in our numerical simulations, as represenatives
of the different regimes we observed. 
Throughout our work, we perform the computation with $M=10$ orbitals to have converged results. We observed negligible quantitative difference between the  computed quantities with $M=10$ and $M=12$ orbitals. Convergence is also assured as the occupation in the last orbital is close to zero.

We observe that for the considered values $g_d= 0,0.002,0.02,0.2$, the ratio between the height of the potential, $V_0$, and
the chemical potential, $\mu$ is, respectively, $3.61, 3.44, 2.52, 0.69$. Since in general the two-mode ansatz is expected to reasonably work
when $V_0 / \mu \gtrsim 1$, then varying $g_d$ we pass from a regime in which $V_0/\mu$ is well larger than $1$ to the one in which smaller than $1$.

The tunneling dynamics is presented through the study of the following quantities: \\

\par
(a) \textit{One-body tunneling dynamics:}\\ From the many-body wavefunction $\psi ( x_1,x_2, \dots ,x_N;t )$,
the reduced one-body density matrix $\rho^{(1)}$ is calculated as
\begin{eqnarray}
\begin{split}
\rho^{(1)}(x^{\prime}\vert x;t)=N\int_{}^{}dx_{2}dx_{3}...dx_{N} \\ \psi^{*}(x^{\prime},x_{2},\dots,x_{N};t)  \psi(x,x_{2},\dots,x_{N};t).
\label{onebodydensity}
\end{split}
\end{eqnarray}
Its diagonal gives the one-body density $\rho (x,t)$, defined as
\begin{eqnarray}
\begin{split}
\rho( x;t)=N\int_{}^{}dx_{2}dx_{3}...dx_{N}  \psi^{*}(x,x_{2},\dots,x_{N};t) \\ \psi(x,x_{2},\dots,x_{N};t).
\label{onebodydensity2}
\end{split}
\end{eqnarray}
It gives the
%probability of a particle
density at the position $x$ and time $t$, when the contribution of other particles is traced out. 

For each choice of $g_d$, we calculate $\rho(x,t)$ till time $t=150$. We present $\rho(x,t)$ in Fig.\ref{Fig1} for different values
of $g_d$. Comparison is made with the non-interacting case. In absence of any interactions ($g_d =0$), the atoms simply exhibit
Rabi oscillation between both wells. For $g_d= 0.002$ (a very small value), we observe tunneling dynamics, with
the frequency and amplitude of oscillations significantly affected by the repulsive tail of non-local
interaction. 
We do the computation for several other number of dipolar bosons, $N=3,4,5$ and observe the same dynamics.
For $g_d = 0.02$, the tunneling dynamics exhibits a kind of self-tapping.
The population in the right well is nearly stationary even for long time. 
The self-trapping at a relatively 
weak interaction appears to be a consequence of long-range repulsive tail of the dipole interaction.  For contact interaction (not shown here), we find the self-trapping behaviour but for much higher interaction strength. We report also
the computation with $N=3,4,5$. We observe clear signature of self-trapping at short time dynamics. However, as expected,
due to quantum fluctuations, the bosons do not remain in the self-trapped state for very long time irrespective of the particle number.
For much higher interaction strength, with $g_d = 0.2$, we observe equilibration dynamics. An
equal number of particles settle in the two wells.\\

\par
(b) \textit{Population imbalance:}\\
We further calculate the population in the right well using 
\begin{equation}
\hspace{18ex}  N_R(t) = \int_{0}^{\infty}  \rho (x,t) dx,
\label{nr}
\end{equation}
and plot $\frac{N_R(t)}{N}$ in Fig.~\ref{Fig2} for the same choice of the values of $g_d$. For $g_d=0$, we again observe pure Rabi oscillations
as expected. For $g_d= 0.002$, as we introduce small correlations, we observe two different scale of dynamics.
Up to a time $t\simeq 80$, the number of bosons in the right well fluctuates. At longer time dynamics, we observe a trend of partial equilibration,
with the value of $\frac{N_R(t)}{N}$ trying to saturate at $0.5$. When $g_d = 0.02$, which is ten times larger than the previous one,
we observe a single scale of dynamics, with $\frac{N_R(t)}{N}$ remaining close to unity throughout the dynamics.
From the direct calculation of the population in the right well, we estimate that for this choice of $g_d$, the self-trapping is close to 90\%.
When we analyze the dynamics with $g_d = 0.2$, the system shows a different type of single scale dynamics i.e. equilibration.
With $N_L(t) = \int_{-\infty}^0 \rho(x,t) dx$, we find in this state $N_L \simeq N_R\simeq 3$.
Thus when the strength of dipolar interaction is tuned, we observe four kinds of dynamical features: pure Rabi oscillation,
deformed Rabi oscillation with some signature of partial equilibration at large times, nonlinear self-trapping and equilibration. \\
\par
We also compute the population imbalance as 
\begin{equation}
\hspace{16ex} Z(t)= \frac{N_R(t)-N_L (t)}{N}
\label{zt}
\end{equation}
and its average value calculated as 
\begin{equation}
\hspace{18ex}  Z_{avg}= \frac{1}{t} \int_0^t Z(t^{\prime}) dt^{\prime}
\label{zt_avg}
\end{equation}

\begin{figure}  %%%fig3 imbalance
	\begin{center}
		\includegraphics[height=.350\textwidth,angle=-90]{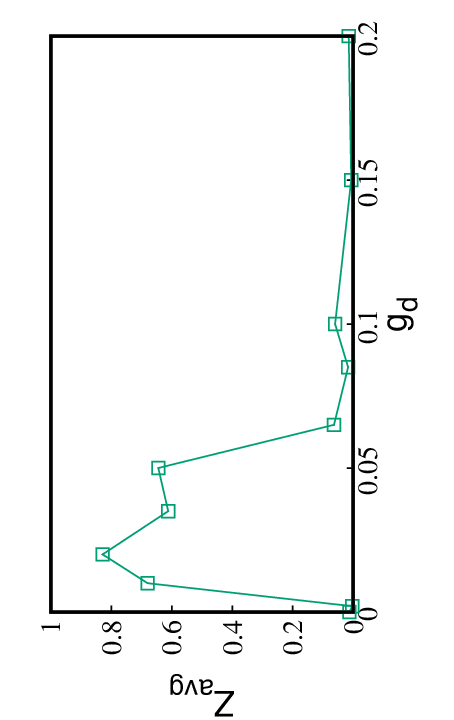}
	\end{center}
	\caption{Time evolution of average value of population imbalance as a function of the interaction strength $g_d$. The peak point corresponds to the maximum self-trapping.}
	\label{Fig3}
\end{figure}

 \begin{figure}  %%%fig4 two-body 0.002
	\begin{center}
		\includegraphics[height=.66\textwidth,angle=0]{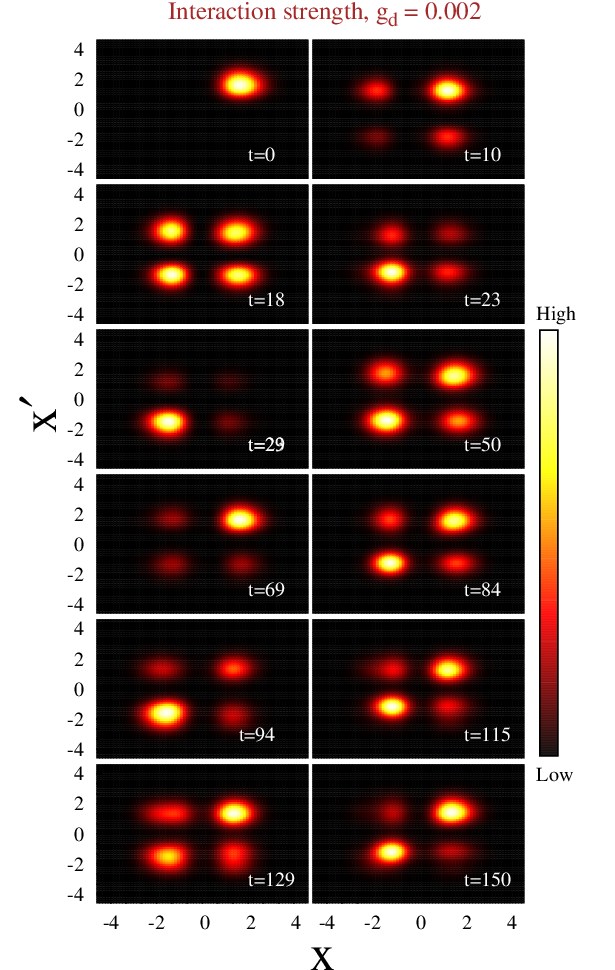}
	\end{center}
	\caption{Time evolution of the two-body density for $g_d=0.002$.}
	\label{Fig4}
\end{figure}

 We plot $Z_{avg}$ as a function of $g_d$ in Fig.~\ref{Fig3}. It exhibits a kind of crossover between the four kind of dynamics reported above.
 $Z_{avg}$ is zero for the two extreme cases, namely Rabi oscillation for non-interacting limit and equilibration dynamics for strong
 interaction. Whereas maximum population imbalance occurs at some interaction strength $g_d^c$ which corresponds to self-trapping.
 Thus the width of the curve qualitatively gives an estimate to have the interaction regime for self-trapping.\\

\par
(c) \textit{Two-body correlation dynamics:}\\
 Next we calculate the time evolution of the two-body density and investigate how the effect of two-body correlation comes in
 the dynamics increasing $g_d$. The second order reduced density matrix is defined as
 \begin{eqnarray}
\begin{split}
\rho^{(2)}(x_{1}^{\prime}, x_{2}^{\prime} \vert x_{1},x_{2};t)=  N(N-1)\int_{}^{}dx_{3}dx_{4}....dx_{N} \\ \psi^{*}(x_{1}^{\prime},x_{2}^{\prime},x_{3},...,x_{N};t) \psi(x_{1},x_{2},...x_{N};t)
\end{split}
\end{eqnarray}
and the diagonal part of the two-body density is given by
\begin{equation}
\rho^{(2)}(x_{1},x_{2};t) \equiv \rho^{(2)}(x_{1}^{\prime}=x_{1},x_{2}^{\prime}=x_{2} \hspace{1ex} \vert \hspace{1ex} x_{1},x_{2};t).
\end{equation} 

\begin{figure}  %%%fig4 two-body 0.02
	\begin{center}
		\includegraphics[height=.45\textwidth,angle=0]{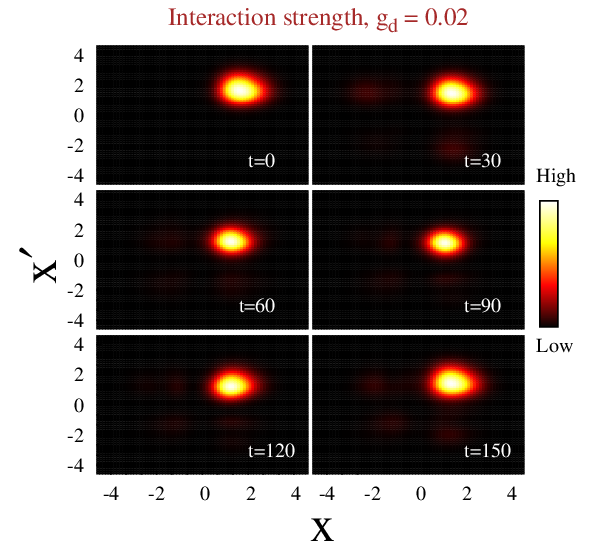}
	\end{center}
	\caption{Time evolution of spatial two-body density for $g_d=0.02$.}
	\label{Fig5}
\end{figure} 

\begin{figure}  %%%fig6 two-body 0.2
	\begin{center}
		\includegraphics[height=.50\textwidth,angle=0]{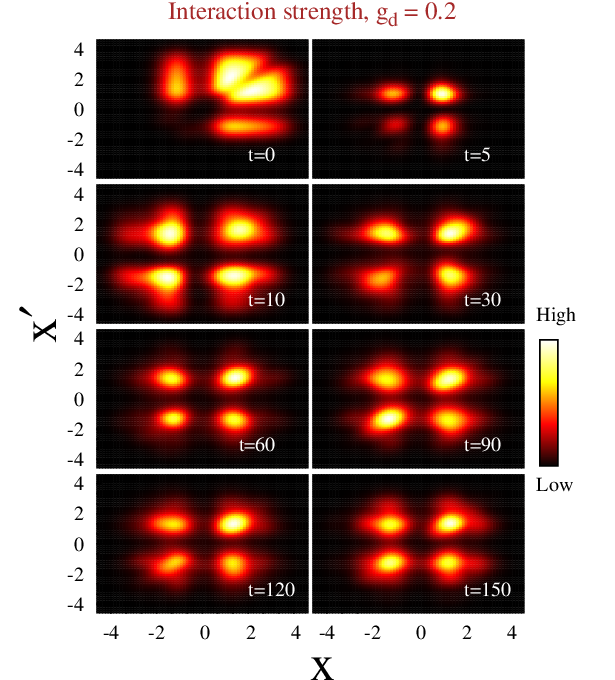}
	\end{center}
	\caption{Time evolution of spatial two-body density for $g_d=0.2$.}
	\label{Fig6}
\end{figure}

Fig.\ref{Fig4} - Fig.\ref{Fig6} plot the results for the spatial two-body density for $g_d = 0.002$, $0.02$ and $0.2$ respectively.
Non-interacting bosons tunnel independently and they are not shown. For inteacting atoms,
the effect of two-body correlation is measured by the competition between the non-local interaction and the inter-well coherence.
Thus the two-body correlation dynamics expected to be complicated.
For larger separation $\vert x-x^{\prime}\vert >> a_\perp$ (characteristic length of transverse confinement),
the interaction potential varies as $W( x-x^{\prime}) \sim \frac{1}{( x-x^{\prime})^3}$ where $\vert x-x^{\prime}\vert \leqslant a_\perp$,
for small separation the transverse confinement induces a short range cutoff $\alpha_0 = a_\perp$. 
Snapshots of two-body density $\rho^{(2)}(x_1,x_2;t)$ for $g_d =0.002$ are presented in Fig.~\ref{Fig4} for different times.
At such a weak interaction, the many-body states are fully coherent, with the six bosons clustered in the right well. 
%We consider $t=18$ as the equilibrium point from Fig.~\ref{Fig2}, when $\frac{N_R(t)}{N} \simeq \frac{N_L(t)}{N} =0.5 $, the inter-well coherence is completely maintained--- atoms are likely to be in both wells. 
However, during the tunneling, we observe competition between the correlation due to non-local interaction and the inter-well coherence.
With evolution of time, due to the repulsive tail in the interaction, the atoms repel each other and the atoms start to tunnel
(snapshot at time $t=10$). At larger times (e.g. $t=18$), we observe four equal bright spots,
two diagonal ($x=x^{\prime}$) bright spots corresponding to equal number of bosons residing in the two wells,
and the other two bright spots along anti-diagonal ($x=-x^{\prime}$) expressing the fact that the inter-well coherence is maintained.
Thus interatomic correlation and inter-well correlations are maintained in the same scale. During the course of tunneling,
we can also verify that at this later time, $\frac{N_R(t)}{N} \simeq \frac{N_L(t)}{N}  \simeq 0.5 $ from Fig.~\ref{Fig2}.
At laters time (e.g. $t=29$), we observe highly coherent cluster set up in the left well, showing the loss of the
inter-well coherence and the prevalence of the interatomic correlations. With time evolving, we observe a rather complex competition
between the interatomic correlations in either well and inter-well coherence, which results in the Rabi-like oscillations
shown in Fig.~\ref{Fig2}. At longer times, we observe only two bright spots along the diagonal which again corresponds to equal population
in two wells ($t=150$). Comparing with the snapshot at time $t=18$, which also corresponds to equal population in two wells, at time $t=150$ the
equal population is obtained at the cost of losing inter-well coherence.  

In Fig.~\ref{Fig5}, we plot the two-body density taking snapshots at different times for $g_d=0.02$. In this case,
the initially coherent cluster settled in the right well remains same all throughout the time which corresponds
to self-trapping as shown in Fig.~\ref{Fig2}. The interatomic correlation plays the crucial role.
However some faded off-diagonal patches appear which signify that exactly 100\% self-trapping is not obtained.
It is in good agreement with Fig.~\ref{Fig2} which exhibits that for $g_d=0.02$, about 90\% self-trapping.
Thus the effect of inter-well coherence is very small.

In Fig.~\ref{Fig6}, we present the two-body density for stronger interaction $g_d=0.2$ at different times. Initially at $t=0$
we observe the strong effect of the tails of the repulsive interaction tails:
the many-body state is incoherent and diffused unlike the case reported in Fig.~\ref{Fig4} and Fig.~\ref{Fig5}. 
%From Fig.~\ref{Fig2}, we have observed that the system quickly attains equilibration and maintain that for long-time. 
In Fig.~\ref{Fig6} we observe that at time $t=10$ four bright spots are developed which corresponds to the equal population in the two wells.
However comparing this with the snapshot at time $t=18$ of Fig.~\ref{Fig4}, we can clearly identify that the bright spots are now diffused
which signifies that three bosons residing in a well feel the effect of the tails of repulsive interaction. Thus atoms are incoherent in each well.
With time, the equilibration is almost maintained, however the interatomic coherence in each well depends on time.
The equilibration dynamics for $g_d=0.2$ is also remarkably different as compared with the case for $g_d=0.002$ (Fig.~\ref{Fig4}).
At time $t=150$ of Fig.~\ref{Fig4}, we observe almost equal population in both wells, but {\textit{absence}} of
inter-well coherence whereas at the same time in Fig.~\ref{Fig6}, equilibration is maintained in {\textit{presence}}
of inter-well coherence. Thus we conclude that the observed dynamics for different choices of interaction strength
is fundamentally governed by the interplay between the interatomic correlation and inter-well coherence.\\

\par
\begin{figure}  %%%fig7 two-body 0.2
	\begin{center}
		\includegraphics[height=.510\textwidth,angle=-90]{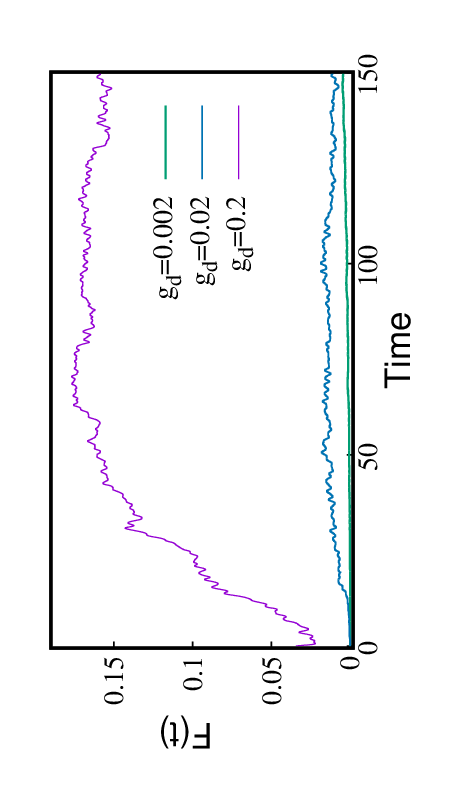}
	\end{center}
	\caption{Plot of $F(t)$ for different interaction strengths. For $g_d=0.002$ and $0.02$, system is less fragmented
          whereas for $g_d=0.2$ system is more fragmented and shows a plateau structure.}
	\label{Fig7}
\end{figure}

 (d) \textit{Fragmentation dynamics:}\\ 
In mean-field theory, one orbital is macroscopically occupied and the system is condensate. However the phenomenon
of fragmentation comes in the picture when more than one orbital is significantly occupied. In our case with $N=6$ particles,
we define a quantfier of the observed fragmentation in the dynamics as 
\begin{equation}
\hspace{16ex} F(t)= 1-\frac{1}{N} \sum_i n_i (t)
\label{frag}
\end{equation} 
where $n_i(t)$ are the time-dependent natural occupations. Thus $F(t)$ quantifies the contribution coming from other significant orbitals.
For $g_d=0.002$, and $0.02$, $F$ is close to zero all throughout the time evolution. However for $g_d=0.2$, even at $t=0$,
the system is initially fragmented which is also seen in  Fig.~\ref{Fig7} ($t=0$). In a very short time, $F$ increases and reaches a constant value.
Then  the plateau is maintained for the rest of the time. The plateau corresponds to the equilibration dynamics when
the bosons equally distributed in the two wells. In the long-time dynamics (not shown here), we observe the plateau nature is
maintained. As $F$ does not increase any more, the occupation in the orbitals remain unchanged:
interatomic correlation does not increase further. There is no competition any more between the inter-well coherence and interatomic correlation:
thus the equilibration dynamics is maintained for very long time.

\section{Conclusion}\label{conclusion}
We studied the few-body quantum
dynamics of few dipolar bosons in a symmetric double-well.
%We report that the dynamics is much richer than that reported for contact interaction.
We studied Rabi oscillations in the non-interacting limit and self-trapping and finally equilibration dynamics for strong interaction.
The complete dynamics for different choices of the interaction strength is presented through the one-body tunneling dynamics.
We further considered the average population imbalance in the two wells as the figure of merit of the tunneling dynamics. The
structure in the average population imbalance qualitatively describe the range of interaction strength to have self-trapping,
as seen in Fig.~\ref{Fig3}. We observed the very interesting dynamics in the two-body density and established the
link between the tunneling dynamics and the role played by the interatomic correlation and inter-well coherence.
The most subtle dynamics is observed at the strong interaction strength when the many-body dynamics exhibits equilibration.
Finally we observed that for equilibration dynamics, the many-body state is initially fragmented and
then exhibits a plateau structure in the long time dynamics.

Several open questions are suggested by the results presented in this paper.
The few-body quantum dynamics gives insights on the phases emerging
from the combined presence of the non-local dipolar interaction and of the
one-dimensionality. Nevertheless, a full study of the quantum dynamics
of dipolar long-range atoms in one-dimensional double-well and the
corresponding diagram of dynamical regimes remain to be explores, and one has
to see how many-body effects in one-dimension modify the regimes holding
for the few-body cases we examined. In partcular,
it would be important to address
how the non-local nature of the interaction modifies the corresponding
results for atoms with short-range interactions in one-dimensional double-well
setups. Moreover, since for very large values of $g_d$, dipolar gases 
exhibits crystallization (see e.g. ~\cite{macri2,sangita}),
the double-well dynamics in this regime would be interesting to study. We also
mention that in the standard quantum two-mode model one can rewrite the density-density interaction term between the populations
of different wells in terms of the local interaction term: therefore, for high barriers, one could ask whether
the tunneling dynamics in double-well potentials can be written in terms of an effective contact interaction and
how the possibility of performing such a mapping depends on the dimensionality of the systems. We finally mention
that the study of the quantum dynamics in presence of both non-local or dipolar and contact interactions in double-well potentials
is also to a deserving subject of future investigation. 

%We show that the dynamics of bosonic Josephson junction for the dipolar bosons is much richer than what is known for contact interaction. \\

% interplay

\begin{acknowledgements}
        R. Roy acknowledges the University Grant Commission (UGC) India for the financial support as a senior research fellow. B. Chakrabarti acknowledges ICTP support where major portion of the work was done.
\end{acknowledgements}


\begin{thebibliography}{200}

\bibitem{5} A. Smerzi,  S. Fantoni,  S. Giovanazzi, and S. R. Shenoy,  Phys. Rev. Lett. \textbf{79}, 4950 (1997).

\bibitem{andrea1} F. S. Cataliotti, S. Burger, C. Fort, P. Maddaloni, F. Minardi, A. Trombettoni, A. Smerzi, and
  M. Inguscio, Science \textbf{293}, 483 (2001).

\bibitem{prl95} M. Albiez, R. Gati, J. F\"olling, S. Hunsmann, M. Cristiani, and M. K. Oberthaler, Phys. Rev. Lett. {\bf 95}, 010402 (2005). 

\bibitem{natphy1} T. Schumm, S. Hofferberth, L. M. Andersson, S. Wildermuth, S. Groth, I. Bar-Joseph, J. Schmiedmayer, and  P. Kr\"uger,
  Nature Phys., \textbf{1}, 57 (2005).
  
\bibitem{andrea2} T. Anker, M. Albiez, R. Gati, S. Hunsmann, B. Eiermann, A. Trombettoni, and M. K. Oberthaler,
  Phys. Rev. Lett. {\bf 94}, 020403 (2005).

\bibitem{Levy} S. Levy, E. Lahoud, I. Shomroni, and J. Steinhauer, Nature \textbf{449}, 579 (2007).

\bibitem{prl98} B. V. Hall, S. Whitlock, R. Anderson, P. Hannaford, and A. I. Sidorov, Phys. Rev. Lett. {\bf 98}, 030402 (2007).
  
\bibitem{nature455} J. Est\`{e}ve, C. Gross, A. Weller, S. Giovanazzi and M. K. Oberthaler, Nature \textbf{455}, 1216 (2008).

\bibitem{LeB}  L. J. LeBlanc, A. B. Bardon, J. McKeever, M. H. T. Extavour, D. Jervis, J. H. Thywissen, F. Piazza, and A. Smerzi,
  Phys. Rev. Lett. \textbf{106}, 025302 (2011).  

\bibitem{contact4} G. Spagnolli, G. Semeghini, L. Masi, G. Ferioli, A. Trenkwalder, S.  Coop, M.  Landini, L.  Pezz\'e,
  G. Modugno, M. Inguscio, A. Smerzi, and M. Fattori, Phys. Rev. Lett. \textbf{118}, 230403 (2017).  

\bibitem{Xhani19} K. Xhani, E. Neri, L. Galantucci, F. Scazza, A. Burchianti, K.-L. Lee, C. F. Barenghi, A. Trombettoni,
  M. Inguscio, M. Zaccanti, G. Roati, and N. P. Proukakis, \verb|arXiv:1905.08893|
  
\bibitem{Liscience} G. Valtolina, A. Burchianti, A. Amico, E. Neri, K. Xhani, J. A. Seman, A. Trombettoni, 
A. Smerzi, M. Zaccanti, M. Inguscio, and G. Roati, Science \textbf{350}, 1505 (2015).

\bibitem{Lidiss} A. Burchianti, F. Scazza, A. Amico, G. Valtolina, 
J. A. Seman, C. Fort, M. Zaccanti, M. Inguscio, and G. Roati, Phys. Rev. Lett. \textbf{120}, 025302 (2018).

\bibitem{Luick} N. Luick, L. Sobirey, M. Bohlen, V. P. Singh, L. Mathey, T. Lompe, and H. Moritz, \verb|arXiv:1908.09776|

\bibitem{Kwon} W. J. Kwon, G. Del Pace, R. Panza, M. Inguscio, W. Zwerger, M. Zaccanti, F. Scazza, and G. Roati, \verb|arXiv:1908.09696|

\bibitem{zapata} I. Zapata, F. Sols, and A. J. Leggett, Phys. Rev. A \textbf{57}, R28 (1998).

\bibitem{rag} S. Raghavan, A. Smerzi, S. Fantoni, and S. R. Shenoy, Phys. Rev. A \textbf{59}, 620 (1999).

\bibitem{st} A. Smerzi and A. Trombettoni, Phys. Rev. A \textbf{68}, 023613 (2003).

\bibitem{ab} D. Ananikian and T. Bergeman, Phys. Rev. A \textbf{73}, 013604 (2006).

\bibitem{cat} D. M. Jezek, P. Capuzzi, and H. M. Cataldo, Phys. Rev. A \textbf{87}, 053625 (2013). 

\bibitem{Jav1997} J. Javanainen and M. Wilkens, Phys. Rev. Lett. \textbf{78}, 4675 (1997).
  
\bibitem{Raghavan1999} S. Raghavan, A. Smerzi, and V. M. Kenkre, Phys. Rev. A \textbf{60}, R1787(R) (1999).

\bibitem{Franzosi2000} R. Franzosi, V. Penna, and R. Zecchina, Int. J. Mod. Phys.
  B \textbf{14}, 943 (2000). 
  
\bibitem{smerzi2000} A. Smerzi and S. Raghavan,
  Phys. Rev. A \textbf{61}, 063601 (2000).

\bibitem{bosehubbard2} R. Gati and M. K. Oberthaler,  J. Phys. B \textbf{40}, R61 (2007).

\bibitem{Giovanazzi} S. Giovanazzi, Est\'eve, and M. K. Oberthaler, New J. Phys. \textbf{10}, 045009 (2008). 

\bibitem{DellAnna} L. Dell'Anna, Phys. Rev. A \textbf{85}, 053608 (2012). 

\bibitem{Bila} M. Bilardello, A. Trombettoni, and A. Bassi, Phys. Rev. A \textbf{95}, 032134 (2017). 

\bibitem{Jo} G.-B. Jo, Y. Shin, S. Will, T. A. Pasquini, M. Saba, W. Ketterle, D. E. Pritchard, M. Vengalattore, and M. Prentiss, 
Phys. Rev. Lett. \textbf{98}, 030407 (2007).

\bibitem{gritsev} V. Gritsev, A. Polkovnikov, and E. Demler,
  Phys. Rev. B \textbf{75}, 174511 (2007).

\bibitem{Aga} K. Agarwal, E. G. Dalla Torre, J. Schmiedmayer, and E. Demler,
  Phys. Rev. B \textbf{95}, 195157 (2017). 
  
\bibitem{aPolo18} J. Polo, V. Ahufinger,
  F. W. J. Hekking, and A. Minguzzi,
  Phys. Rev. Lett. \textbf{121}, 090404 (2018).


\bibitem{Pigneur18} M. Pigneur, T. Berrada, M. Bonneau, T. Schumm, E. Demler, and J. Schmiedmayer, Phys. Rev. Lett. \textbf{120}, 173601 (2018). 

\bibitem{Polo} J. Polo, R. Dubessy, P. Pedri, H. Perrin, and A. Minguzzi,
  Phys. Rev. Lett. \textbf{123}, 195301 (2019).

\bibitem{25} A. I. Streltsov, O. E. Alon, and L. S. Cederbaum,
  Phys. Rev. A \textbf{73}, 063626 (2006).
  
\bibitem{36} A. I. Streltsov, Phys. Rev. A \textbf{88}, 041602(R) (2013).

\bibitem{s40} O. I. Streltsov, O. E. Alon, L. S. Cederbaum, and A. I. Streltsov, Phys. Rev. A \textbf{89}, 061602(R) (2014).
  
\bibitem{fischer15} U. R. Fischer, A. U. J. Lode and B. Chatterjee, Phys. Rev.A \textbf{91}, 063621 (2015).

% Double well, 1D, numerical solution.
  
\bibitem{prl100} S. Z\"ollner, H. Meyer, and P. Schmelcher, Phys. Rev. Lett. {\bf 100}, 040401 (2008). 

\bibitem{contact2} K. Sakmann, A. I. Streltsov, O. E. Alon and L. S. Cederbaum, Phys. Rev. Lett. \textbf{103}, 220601 (2009).

\bibitem{manybody1} K. Sakmann, A. I. Streltsov, O. E. Alon and L. S. Cederbaum,Phys. Rev. A \textbf{82}, 013620 (2010). 

\bibitem{contact3}  K. Sakmann, A. I. Streltsov, O. E. Alon and L. S. Cederbaum,Phys. Rev. A \textbf{89}, 023602 (2014).

\bibitem{Sudip2019} S. K. Haldar and O. E. Alon, New J. Phys. \textbf{21}, 103037 (2019).  

% ---------
  
\bibitem{chromium1} A. Griesmaier, J. Werner, S. Hensler, J. Stuhler, and T. Pfau, Phys. Rev. Lett. \textbf{94}, 160401 (2005).

\bibitem{chromium2} Q. Beaufils, R. Chicireanu, T. Zanon, B. Laburthe-Tolra, E. Mar$\acute{e}$chal, L. Vernac, J.-C. Keller, and O. Gorceix,
  Phys. Rev. A \textbf{77}, 061601 (2008).

\bibitem{dysprosium} M. Lu, N. Q. Burdick, S. H. Youn, and B. L. Lev, Phys. Rev. Lett. \textbf{107}, 190401 (2011).

\bibitem{erbium} K. Aikawa, A. Frisch, M. Mark, S. Baier, A. Rietzler, R. Grimm, and F. Ferlaino, Phys. Rev. Lett. \textbf{108}, 210401 (2012).

\bibitem{30} M. A. Baranov, Phys. Rep. \textbf{464}, 71 (2008).

\bibitem{31} T. Lahaye, C Menotti, L Santos, M Lewenstein and T Pfau, Rep. Prog. Phys. \textbf{72}, 126401 (2009).

\bibitem{tang19} Y. Tang, W. Kao, K.-Y. Li, S. Seo, K. Mallayya, M. Rigol,
  S. Gopalakrishnan, and B. Lev, Phys. Rev. X \textbf{8}, 021030 (2018).
  
\bibitem{vodola14} D. Vodola, L. Lepori, E. Ercolessi, A. V. Gorshkov, and G. Pupillo,
  Phys. Rev. Lett. \textbf{113}, 156402 (2014).

\bibitem{gong16} Z.-X. Gong, M. F. Maghrebi, A. Hu, M. Foss-Feig, P. Richerme,
  C. Monroe, and A. V. Gorshkov, Phys. Rev. B \textbf{93}, 205115 (2016).

\bibitem{lepori16} L. Lepori, D. Vodola, G. Pupillo, G. Gori, and A. Trombettoni,
  Ann. Physics \textbf{374}, 35 (2016).

\bibitem{celardo16} G. L. Celardo, R. Kaiser, and F. Borgonovi, Phys. Rev. B \textbf{94}, 144206 (2016).  

\bibitem{defenu16} N. Defenu, A. Trombettoni, and S. Ruffo, Phys. Rev. B \textbf{94}, 224411 (2016);
  {\em ibid.} Phys. Rev. B \textbf{96}, 104432 (2017).

\bibitem{lepori17} L. Lepori, A. Trombettoni, and D. Vodola, J. Stat. Mech. 033102 (2017).

\bibitem{igloi18} F. Igl\'oi, B. Bla$\beta$, G. Ro\'osz, and H. Rieger,
  Phys. Rev. B \textbf{98}, 184415 (2018).

\bibitem{blass18} B. Bla$\beta$, H. Rieger, G. Ro\'osz, and F. Igl\'oi, Phys. Rev. Lett. \textbf{121}, 095301 (2018). 

\bibitem{defenu18} N. Defenu, T. Enss, M. Kastner, and G. Morigi,
  Phys. Rev. Lett. \textbf{121}, 240403 (2018).

\bibitem{lerose19} L. Lerose, B. Zunkovic, A. Silva, and A. Gambassi,
  Phys. Rev. B \textbf{99}, 121112 (2019).

\bibitem{39} L. Santos, G. V. Shlyapnikov, P. Zoller and M. Lewenstein, Phys. Rev. Lett. \textbf{85}, 1791 (2000).

\bibitem{Sinha} S. Sinha and L. Santos, Phys. Rev. Lett. \textbf{99}, 140406 (2007).
  
\bibitem{Citro} R. Citro, E. Orignac, S. De Palo, and M.-L. Chiofalo, Phys. Rev. A \textbf{75}, 051602 (2007). 
  
\bibitem{35} P. Bader and U. R. Fischer, Phys. Rev. Lett. \textbf{103}, 060402 (2009).

\bibitem{38} N. Henkel, F. Cinti, P. Jain, G. Pupillo and T. Pohl, Phys. Rev. Lett. \textbf{108}, 265301 (2012).

\bibitem{Cinti} F. Cinti, T. Macr\`{i}, W. Lechner, G. Pupillo, and T. Pohl, Nature Communications \textbf{5}, 3235 (2014). 

\bibitem{DeP} S. De Palo, R. Citro, and E. Orignac, Phys. Rev. B \textbf{101}, 045102 (2020). 

\bibitem{abad1} M. Abad, M. Guilleumas, R. Mayol, M. Pi, and D. M.
Jezek, Phys. Rev. A \textbf{84}, 035601 (2011).

\bibitem{abad2} M. Abad, M. Guilleumas, R. Mayol, M. Pi and D. M.
Jezek, Euro. Phys. Lett., \textbf{94} 10004 (2011).

\bibitem{abad3} A. Gallem, M. Guilleumas, R. Mayol, and A. Sanpera, Phys. Rev. A \textbf{88}, 063645 (2013).
  
\bibitem{blume} M. Asad-uz-Zaman and D. Blume, Phys. Rev. A \textbf{80}, 053622 (2009).

\bibitem{mazzarella}  G. Mazzarella and L. Dell'Anna, EPJ ST \textbf{217}, 197 (2013). 
  
\bibitem{halder} S. K. Haldar and O. E. Alon, Chem. Phys. \textbf{509}, 72 (2018).

\bibitem{Sudip2019conf} S. K. Haldar and O. E. Alon, J. Phys.: Conf. Series \textbf{1206}, 012010 (2019). 
  
\bibitem{Olshanii:98} M. Olshanii, Phys. Rev. Lett. {\bf 81}, 938 (1998).

\bibitem{Leggett} A. J. Leggett, Rev. Mod. Phys. \textbf{73}, 307 (2001).  

  
\bibitem{Streltsov2007} A. I. Streltsov, O. E. Alon, and L. S. Cederbaum,  Phys. Rev. Lett. \textbf{99}, 030402 (2007).
									
\bibitem{Ofir2008} O. E. Alon, A. I. Streltsov, and L. S. Cederbaum, Phys. Rev. A \textbf{77}, 033613 (2008).

\bibitem{x} E. Fasshauer and A. U. J. Lode, Phys. Rev. A \textbf{93}, 033635 (2016).

\bibitem{y} Rui Lin et. al. Quantum Sci. Technol. \textbf{5} 024004 (2020).

\bibitem{axelN3} A. U. J. Lode, M. C. Tsatsos, E. Fasshauer, R. Lin, L. Papariello, P. Molignini, C. L\'ev\^eque, and S. E. Weiner,
  \textsc{MCTDH-X}:{\em The time-dependent multiconfigurational Hartree for indistinguishable particles software}, {http://ultracold.org} (2018).
  
\bibitem{axelN2} A. U. J. Lode, Phys. Rev. A \textbf{93}, 063601 (2016).

\bibitem{budha1} B. Chatterjee, C. L\'ev\^eque, J. Schmiedmayer, and A. U. J. Lode Phys. Rev. Lett. \textbf{125}, 093602 (2020).

\bibitem{budha2} Budhaditya Chatterjee and Axel U. J. Lode Phys. Rev. A \textbf{98}, 053624 (2018).

\bibitem{budha3} Budhaditya Chatterjee et. al., New J. Phys. \textbf{21}, 033030 (2019).
  
  

\bibitem{MCTDHB_OCT} J. Grond, J. Schmiedmayer, and U. Hohenester, Phys. Rev. A \textbf{79}, 021603(R) (2009).

\bibitem{MCTDHB_Shapiro} J. Grond, T. Betz, U. Hohenester, N. J. Mauser, J. Schmiedmayer, and T. Schumm, New J. Phys. \textbf{13}, 065026 (2011).
    
\bibitem{cao1} L. Cao, S. Kr\"onke, O. Vendrell, and P. Schmelcher, J. Chem. Phys. \textbf{139}, 134103 (2013).

\bibitem{cao2} R. Schmitz, S. Kr\"onke, L. Cao, and P. Schmelcher, Phys. Rev. A \textbf{88}, 043601 (2013).

\bibitem{NJP2015Schmelcher} J. M. Schurer, A. Negretti, and P. Schmelcher,  New J. Phys. \textbf{17}, 083024 (2015).

\bibitem{axelN6} A. U. J. Lode, B. Chakrabarti, and V. K. B. Kota, Phys. Rev. A \textbf{92}, 033622 (2015).
  
\bibitem{Axel2017} A. U. J. Lode and C. Bruder, Phys. Rev. Lett. \textbf{118}, 013603 (2017).

\bibitem{NJP2017Camille} C. L\'{e}v\^{e}que and L. B. Madsen, New J. Phys. \textbf{19},  043007 (2017). 

\bibitem{NJP2017Schmelcher} G. C. Katsimiga, S. I. Mistakidis, G. M. Koutentakis, P. G. Kevrekidis, and P. Schmelcher,
  New J. Phys. \textbf{19}, 123012 (2017).
  
\bibitem{rhombik} R. Roy, A. Gammal, M. C. Tsatsos, B. Chatterjee, B. Chakrabarti, and A. U. J. Lode, Phys. Rev. A {\bf{97}}, 043625 (2018).

\bibitem{rhombik2} R. Roy, C. L\'{e}v\^{e}que, A. U. J. Lode, A. Gammal, and B. Chakrabarti, Quantum Reports  \textbf{1}, 304 (2019).

\bibitem{rhombik3} S. Bera, R. Roy, A. Gammal, B. Chakrabarti, and B. Chatterjee, J. Phys. B \textbf{52}, 21 (2019).


\bibitem{axelprx} J.H.V. Nguyen, M.C. Tsatsos, D. Luo, A.U.J. Lode,
G.D. Telles, V.S. Bagnato, and R.G. Hulet Phys. Rev.
X \textbf{9}, 011052 (2019).
							 
\bibitem{Ofir2019Rev} A. Lode, C. L\'{e}v\^{e}que, L. Madsen, A. Streltsov, and O. E. Alon, rev. Mod. Phys., \textbf{92} 011001 (2020).
  

\bibitem{LF1} P. Kramer and M. Saracento, \textit{Geometry of the time-dependent variational principle} (Springer, Berlin, 1981).

\bibitem{LF2} H.-J. Kull and D. Pfirsch, Phys. Rev. E \textbf{61}, 5940 (2000).

%\bibitem{4} G. J. Milburn, J. Corney, E. M. Wright and D. F. Walls, Phys. Rev. A \textbf{55}, 4318 (1997).

\bibitem{macri2} T. Macr\`{i}, F. Maucher, F. Cinti, and T. Pohl, Phys. Rev. A \textbf{87}, 061602(R) (2013).
  
\bibitem{sangita} S. Bera, B. Chakrabarti, A. Gammal, M. C. Tsatsos, M. L. Lekala, B. Chatterjee, C. L\'ev\^eque, and A. U. J. Lode,
  Sci. Rep. \textbf{9}, 17873 (2019).

\end{thebibliography}
\end{document}